\documentclass[a4paper]{jpconf}
\usepackage{graphicx}
\usepackage{caption}
\usepackage{subcaption}
\RequirePackage{lineno}
\usepackage{wrapfig}
\usepackage{sidecap}

\usepackage{hyperref}
\usepackage{bookmark}

\begin{document}
\title{Search for muonic atoms at RHIC}

\author{Kefeng Xin}

\address{Physics \& Astronomy Department, Rice University, Houston, Texas 77005}

\ead{kefeng.xin@rice.edu}

\begin{abstract}
We report the search results of muonic atoms from the STAR experiment at RHIC. 
The invariant mass distributions of hadron-muon pairs show peaks at the expected atom masses. 
Two particle correlation functions indicate the hadrons and muons that are close in phase space are emitted at the same space time point, which is a possible signature of muonic atom ionization at the beam pipe. 
The pion-muon correlation functions are further decomposed into two contributions, and the fraction of direct muons is obtained. 
\end{abstract}

\section{Introduction}
Muonic atoms are exotic atoms made of hadrons and muons bounded with the Coulomb force. 
In such atoms, the electrons in ordinary atoms are replaced by muons. 
The simplest muonic atom, made of a proton and a muon, also known as muonic hydrogen, 
has been widely used in precise measurement of the proton structure \cite{protonSize} . 
The $\pi$-$\mu$ muonic atom has been produced from $K_L^0$ decay from very intense beams \cite{pimuBNL,pimuFermi}. 
However, muonic atoms with more exotic cores, such as anti-matter muonic hydrogens and muonic atoms with strangeness have never been observed. 
The high-energy heavy-ion experiments provide a high multiplicity environment of hadrons and leptons. 
It is predicted that right after the collisions, the produced hadrons and muons can be bound together by the Coulomb force and form muonic atoms \cite{baym}. 

In heavy-ion experiments, leptons are considered a good probe of the quark-gluon plasma (QGP) by providing measurement of the electromagnetic emission from the medium without strong interaction in the later stages. 
However, these early produced leptons are mixed with leptons from hadronic weak decays. 
The later, undergoing complex processes, carry indirect information of the hot and dense matter and thus complicate the study of the lepton probe. 
Using muonic atoms addresses this issue as hadronic weak decay muons mostly are produced relatively late in the evolution of the heavy-ion collision without much opportunity to bind with hadrons into atomic systems \cite{baym,kapusta}. 
Therefore the muonic atoms are believed to be an ideal tool to explore the QGP properties. 

\section{Data Analysis and Results}
This study used the dataset collected by the STAR experiment at RHIC at $\sqrt{s_{NN}}=$ 200 GeV from Au+Au collisions in year 2010. 
About 231 million central triggered events passed the event selections and were used in this analysis. 
Two main detectors were used for particle identification: the Time-Projection Chamber (TPC) and the Time-of-Flight detector (TOF). 
The energy loss measured from the TPC and timing information measured from the TOF are combined to select a very pure muon sample between 0.15-0.25 GeV/c. 
The corresponding kaons and protons which fall into the momentum ranges 0.70-1.17 GeV/c and 1.33-2.22 GeV/c can also be cleanly identified. 

\subsection{Invariant Mass}

With the proper particle identification, the atom invariant mass distributions can be studied. 
Three combinatorial methods were used, the unlike-sign, the like-sign, and the mixed-event. 
The like-sign pair mass was corrected from the acceptance difference between positive and negative charged particles:
\begin{eqnarray}
LS_{+-}(corrected)=\sqrt{LS_{++}LS_{--}}\frac{ME_{+-}}{\sqrt{{ME_{++}ME_{--}}}},
\end{eqnarray}
where $LS$ stands for like-sign, $ME$ for mixed-event, and the indices of each term stand for the electric charges for hadrons and leptons. 
The origin of the correction is the different bending direction of particles with different charges in the magnetic field inside the detector. 
The consequence is that pairs with certain opening angles are lost at the TPC sector boundaries or dead TPC readout units. 
The opening angles are different between same-charge and opposite-charge pairs, and lead to lost pairs in different mass regions. 
More discussion of this correction can be found in \cite{dielectron}.
\begin{figure}[!hbt]
	\begin{subfigure}[b]{0.45\textwidth}
		\includegraphics[width=\textwidth]{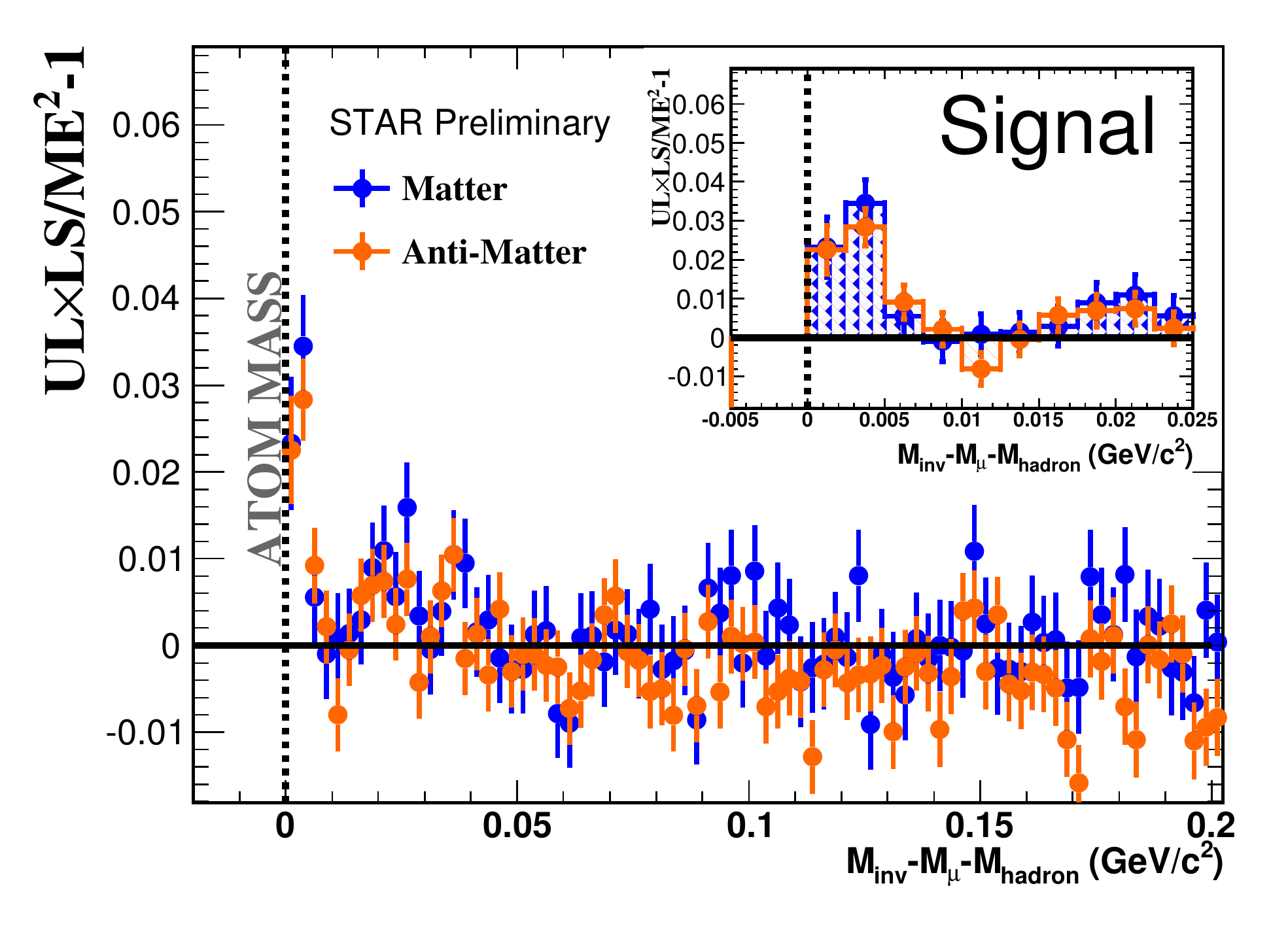}
		\caption{$p$-$\mu^-$ and $\bar{p}$-$\mu^+$ pairs.}
	\end{subfigure}
	\begin{subfigure}[b]{0.45\textwidth}
		\includegraphics[width=\textwidth]{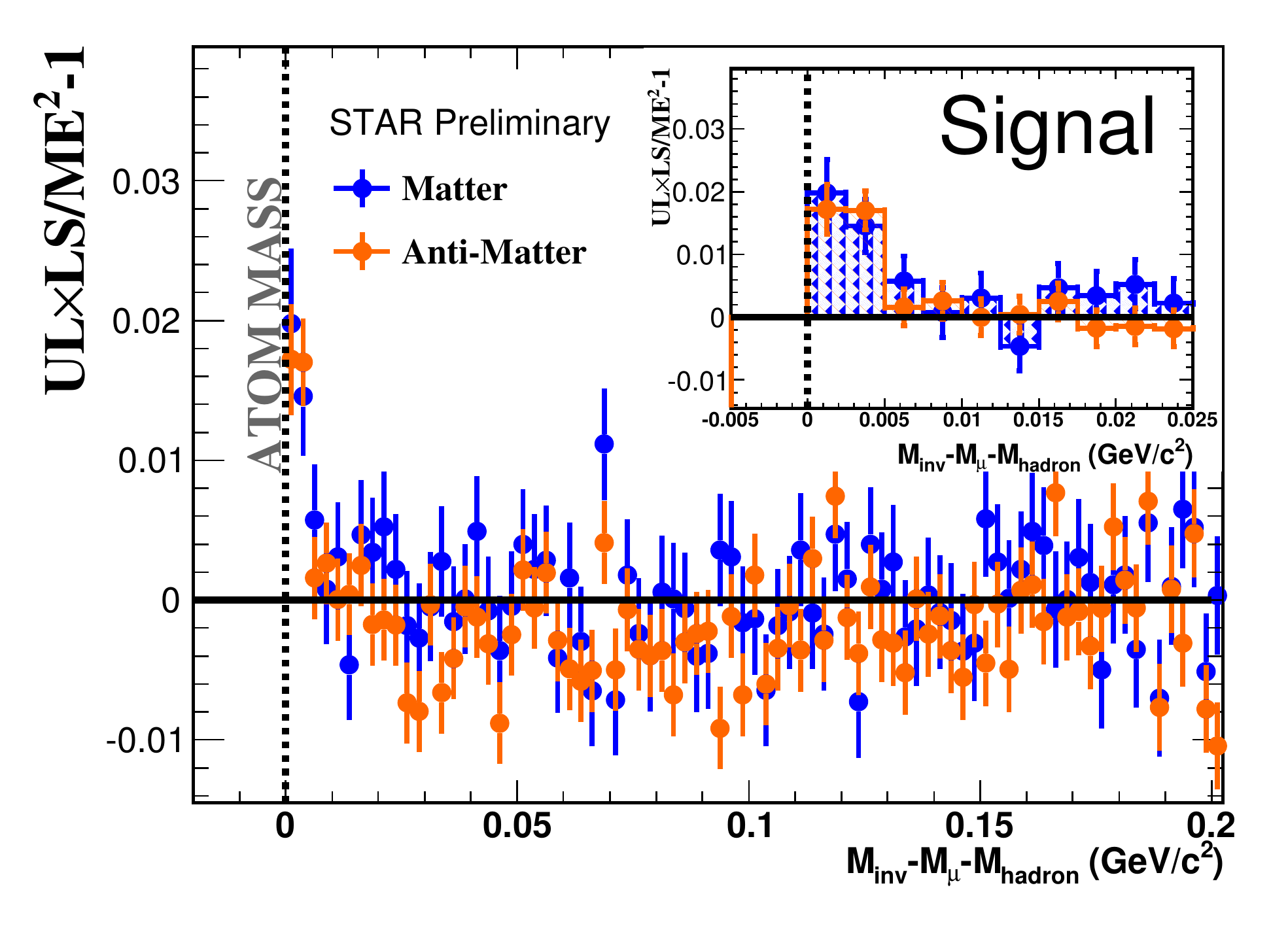}
		\caption{$K^-$-$\mu^+$ and $K^+$-$\mu^-$ pairs.}
	\end{subfigure}
	\caption{The pair invariant mass distributions of $UL\times LS/ME^2-1$ show peaks at the atom masses.}
	\label{fig:mass}
\end{figure}

The unlike-sign method, which pairs a hadron and a muon with opposite electric charges, contains the atom signal, a combinatorial background, and the attractive Coulomb effect. 
The like-sign method, which pairs a hadron and a muon with the same change, contains a combinatorial background and repulsive Coulomb effect. 
The mixed-event method, which pairs a hadron and a muon with opposite charges from different events, contains only a combinatorial background. 
Note that the attractive Coulomb force in the unlike-sign method will enhance the distributions; 
and the repulsive Coulomb force in the like-sign method will suppress the distributions. 
The following variable was used to cancel the Coulomb effect in like-sign and unlike-sign and reveals the correct signal:
\begin{eqnarray}
UL\times LS/ME^2-1,
\end{eqnarray}
where $UL\times LS$ stands for unlike-sign $\times$ like-sign, which cancels the Coulomb effect, and ME stands for mixed-event for normalization. 
Figure\ \ref{fig:mass} shows sharp peaks at zero net mass in $UL\times LS/ME^2-1$ distributions after the Coulomb effect was removed. Here the net mass is the atom mass subtracting from hadron and muon masses.

\subsection{Femtoscopic Correlations}
Two particle correlations have been used to study the interactions between the two daughter particles. 
The correlation function $C(k^*)$, where $k^*$ is the magnitude of momentum in the pair rest frame, describes the strength of the correlation. 
A typical two-particle correlation function shows large deviations from unity if there are only Coulomb interactions \cite{femto}. 
For non-identical particles, a particular particle species can be chosen as a leading particle, and two scenarios can be discussed. 
The first scenario is that the leading particle is produced on average closer to the center of the collision.
If it travels faster than the other particle, the leading particle tends to catch up with the other particle. 
As their distance gets smaller, the correlation, denoted as $C_+$, gets stronger because of more interactions. 
If the leading particle travels slower than the other particle, it tends to move relatively away from the other particle. 
Large distance then leads to weaker interactions and correlations, noted as $C_-$. 
The double ratios $C_+/C_-$ in this scenario deviate from unity, and fall on the same side as the correlation function with respect to unity. 
In the second scenario, the leading particle is produced further away from the center of the collision than the other particle. 
In this scenario, following the similar logic, $C_+/C_-$ also deviates from unity. 
The double ratio was used as an indication of the space time asymmetry of the production of the two particles. 
\begin{figure}[h]
\includegraphics[width=0.48\textwidth]{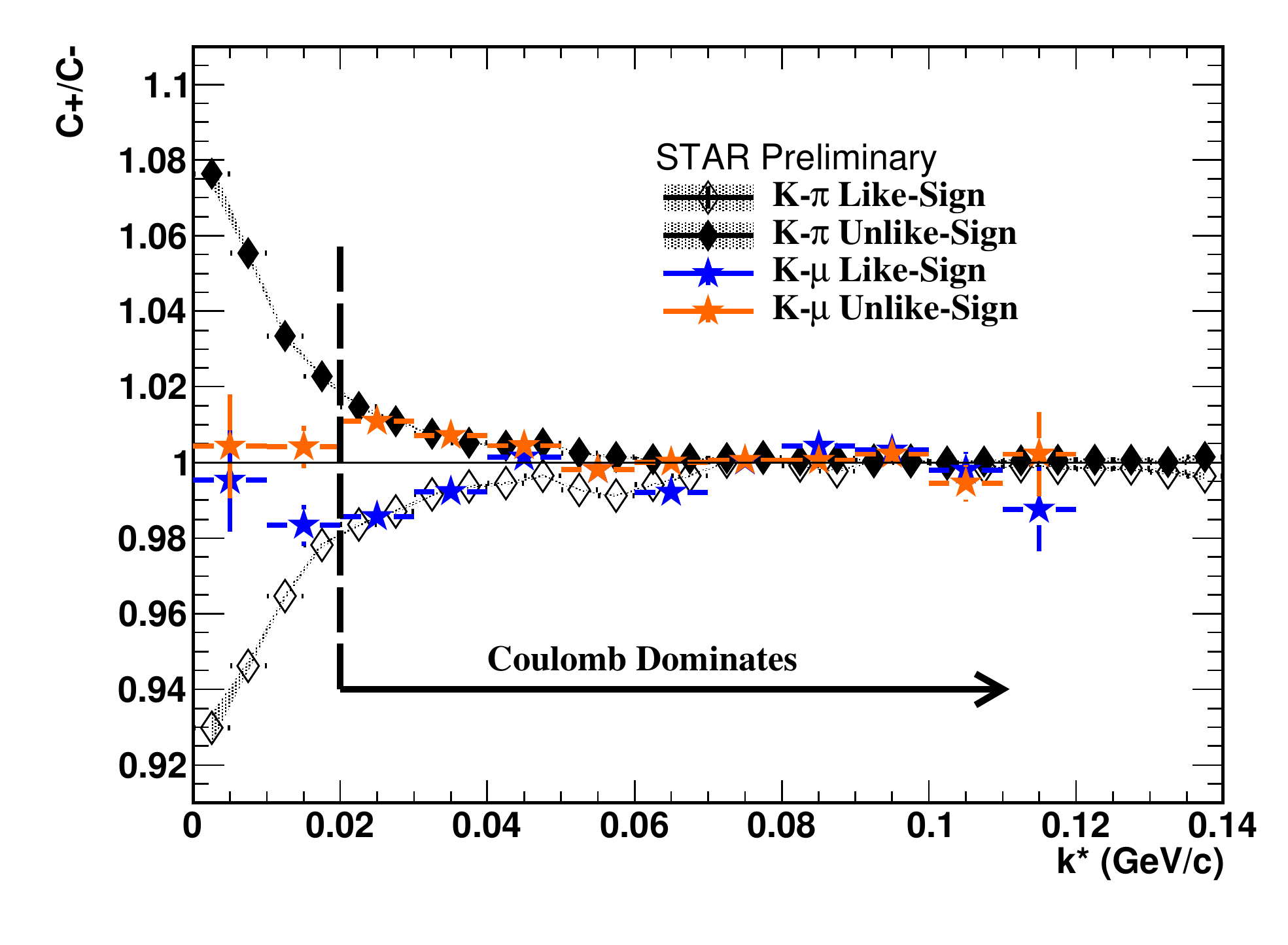}\hspace{2pc}%
\begin{minipage}[b]{14pc}\caption{\label{fig:k4}The double ratio of the $K$-$\pi$ and $K$-$\mu$ systems show significant difference at low $k^*$. The convergence to unity of $K$-$\mu$ suggests atom ionization at the beam pipe, which is a beryllium foil and was used to disassociate the hadron and the muon in an atom.}
\end{minipage}
\end{figure}

The $K$-$\pi$ system has been thoroughly studied \cite{femto} and we calculated the correlation functions as a reference. 
The pion momenta are selected to match the muons. 
We observed the double ratio $C_+/C_-$ in $K$-$\pi$ system deviates from unity as shown in Fig.\ \ref{fig:k4}. 
The smaller the $k^*$ is, the more deviation from unity. 
This monotonic behavior comes from the Coulomb effect, and is an indication of the space-time asymmetry of the production of the pions and the kaons. 
Compared with the reference system, the $K$-$\mu$ system can be divided into two regions indicated in Fig.\ \ref{fig:k4} by the dashed line. 
The Coulomb effect dominates at high $k^*$, and we observed consistency between $K$-$\pi$ and $K$-$\mu$ systems.
At very low $k^*$, where we expect the signal from atoms to appear, we observe significant differences between the two systems. 
The double ratio $C_+/C_-$ in $K$-$\mu$ system converges to unity instead of diverging. 
This is an indication that the two particles are emitted at the same space-time point, 
which is consistent with the dissociation of the hadron and muon at the detector beam pipe. 
The correlation study provides a signature of muonic atom ionization when the atoms hit the detector beam pipe. 

\subsection{$\pi$-$\mu$ Correlations}

\begin{figure}[h]
\includegraphics[width=0.48\textwidth]{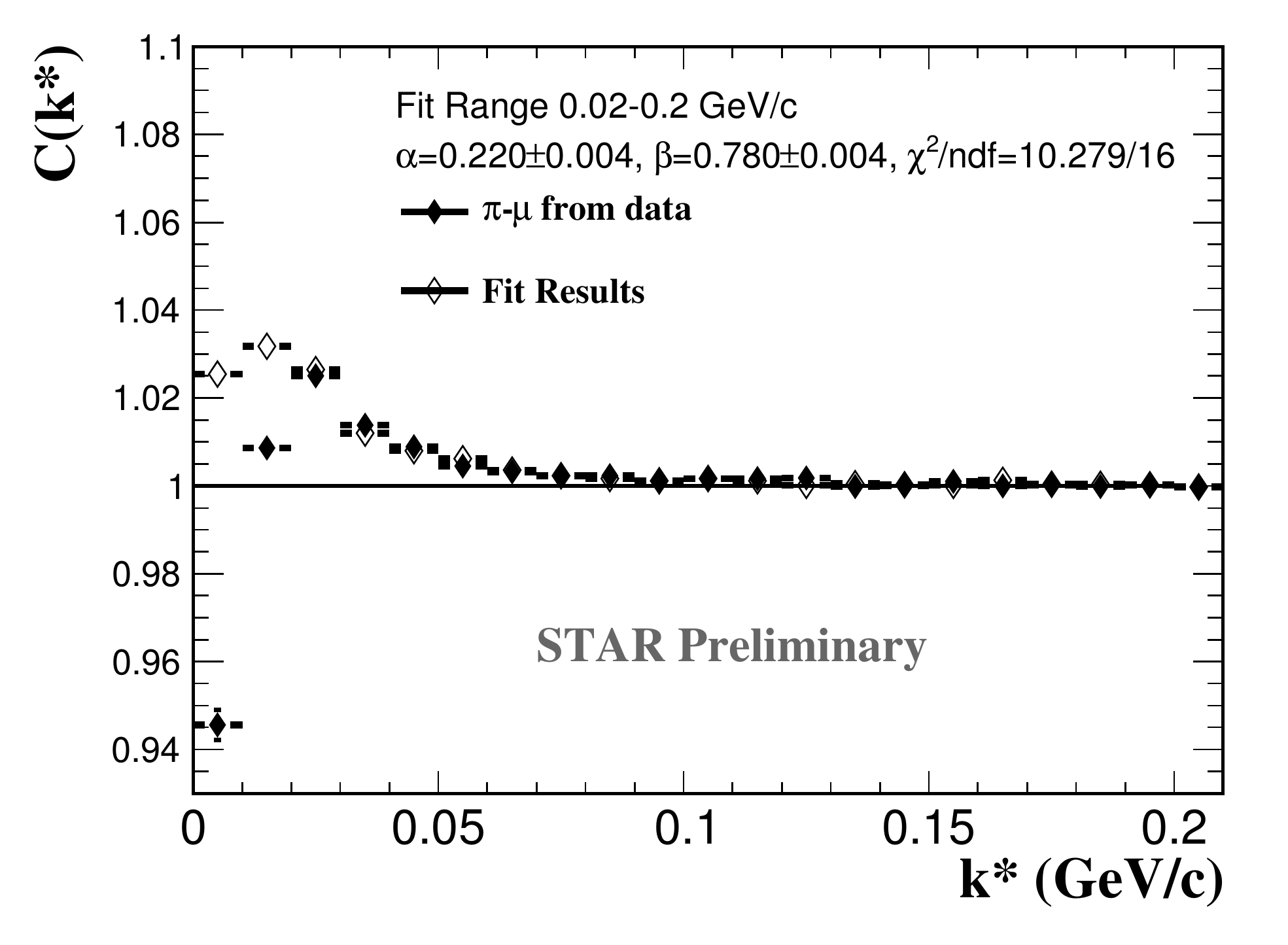}\hspace{2pc}%
\begin{minipage}[b]{14pc}\caption{\label{fig:fraction}Measured $\pi$-$\mu$ correlation function, fitted by $\pi$-$\pi$ correlation function and simulated $\pi$-$\mu_\mathrm{decay}$. 
  The coefficient $\alpha$ represents the weight of the contribution from direct muon and pion correlation. }
\end{minipage}
\end{figure}

The $\pi$-$\mu$ correlations were used to extract the fraction of direct muons. 
The measured correlation functions, denoted as $B$, have two major contributions, the correlation from directly produced muons and pions, and the correlation from weak decay muons and pions. 
The later one inherits the correlations from two pions. 
The two pion correlation, besides the Coulomb effect, has large quantum interference effect from identical particles, which appears to be an attractive force and enhances the correlation \cite{pionHBT}. 
We estimated the first contribution from $\pi$-$\pi$ correlations, considering the pion mass is relatively close to muon mass. 
And we used the reversed unlike-sign $\pi$-$\pi$ correlation to avoid the quantum effect, noted as $1/C$. 
The weak decay muon and pion correlation function was obtained with simulation technique. 
One of the two pions taken from the data is manually set to decay to a muon and a neutrino based on energy momentum conservation. 
Then the correlation function between this artificial muon and the other pion is calculated, denoted as $A$. 
A linear relationship is established: $B=\alpha \times 1/C +\beta \times A$, 
where the coefficient $\alpha$ stands for the fraction of direct muons. 
A $\chi^2$ minimization is then performed to extract the parameters. 
Note that when two pions have very similar momentum, the helices that they produce in the TPC will overlap and cannot be separated. 
The consequence is that one pion is missing in data and cannot be saved. 
The lost pions caused lost muons at very low $k^*$. 
So in the fitting, the low $k^*$ region was discarded, and the fitting range is indicated on Fig.\ \ref{fig:fraction}. 
The fitting results show $22.0 \pm 0.4 \%$ of the muons out of the inclusive muons are direct muons.

\section{Summary}
We presented the first search results on antimatter muonic atoms and $K$-$\mu$ atoms from Au+Au collisions at $\sqrt{s_{NN}}=200$ GeV at the STAR experiment. 
The invariant mass distributions show sharp peaks at the expected mass positions. 
The peaks are consistent across different combinations of hadrons and muons. 
The correlation functions show a signature of muonic atom ionization at the beam pipe, indicating the two constituent particles are disassociated at the same space and time. 
The two components of the $\pi$-$\mu$ correlation function were extracted and the fraction of direct muons out of all produced muons is found to be $22.0\pm 0.4\%$.


\section*{References}
\begin{thebibliography}{9}
\bibitem{protonSize} A. Antognin, {\it et~al.}, Science {\bf 339} 417 (2013); Y. Tanaka and B. M. Steffen, Phys. Rev. Lett. {\bf 51} 18 (1983).
\bibitem{pimuBNL} R. Coombes, {\it et~al.}, Phys. Rev. Lett. {\bf 37} 5 (1976).
\bibitem{pimuFermi} S. H. Aronson, {\it et~al.}, Phys. Rev. Lett. {\bf 48} 16 (1982).
\bibitem{baym} G. Baym, G. Freidman, R. J. Hughes, and B. Jack, Phys. Rev. {\bf D48} 9 (1993).
\bibitem{kapusta} J. Kapusta and A. Mocsy, Phys. Rev. C {\bf59} 5 (1999).
\bibitem{tpc} M. Anderson,\ {\it et al.}, Nucl. Instrum. Methods Phys. Res., Sect. A, Accel. Spectrom. Detect. Assoc. Equip., 499 (2003), p. 659.
\bibitem{tof} J. Adams, {\it et al.}, Phys. Rev. Lett. 94 (2005) 062301; J. Adams, {\it et al.}, Phys. Lett. B 616 (2005) 8.
\bibitem{dielectron} A. Adare, {\it et~al.}, Phys. Rev. C {\bf81} 5 (2010); L. Adamczyk, {\it et~al.}, Phys. Rev. C {\bf86} 5 (2012).
\bibitem{femto} J. Adams, {\it et~al.}, Phys. Rev. Lett. {\bf 91} 26 (2003).
\bibitem{pionHBT} J. Adams, {\it et~al.}, Phys. Rev. C {\bf 71} 044906 (2005).
\end{thebibliography}
\end{document}